\documentclass[structabstract]{aa}  
\usepackage{graphicx}
\usepackage{txfonts}
\usepackage{lscape}
\begin{document}
   \title{Optical linear polarization in ultra cool dwarfs}

   \subtitle{A tool to probe dust in the ultra cool dwarf atmospheres}

   \author{R. Tata
          \inst{1}
          \ ,
          E. L. Mart\'in\inst{2} \inst{3} \inst{1}
          \ ,
          S. Sengupta\inst{4} \inst{5}
          \ ,
          N. Phan-Bao\inst{6}
          \ ,
          M. R. Zapatero Osorio\inst{2} \inst{3}
          \ ,
          H. Bouy\inst{3}
          }
   \institute{University of Central Florida, Department of Physics, Orlando, FL 32816-2385, USA 
              \email{tata@physics.ucf.edu}
         \and 
          Centro de Astrobiolo\'iga (CAB-CSIC), Ctra. Ajalvir km 4, 28850 Torrej\'on de Ardoz, Madrid, Spain
         \and
             Instituto de Astrof\'isica de Canarias,38205 La Laguna,Tenerife,Spain 
          \and         
                 Indian Institute of Astrophysics,Koramangala,Bangalore,560034 India 
          \and
           TIARA-ASIAA/National Tsing Hua University, Hsinchu, Taiwan
           \and
          Institute of Astronomy and Astrophysics, Academia Sinica, P.O. Box 23-141, Taipei 106, Taiwan, China
  }
              

 
  \abstract
   {}
   {Recent studies have detected linear polarization in L dwarfs in the optical I band. Theoretical models
   have  been developed to explain this polarization. These models predict higher polarization at shorter wavelengths. We discuss the polarization in the R and I band of 4 ultra cool dwarfs.}
   {We report linear polarization measurements of 4 ultra cool dwarfs in the R and I bands using the Intermediate dispersion Spectrograph and Imaging System (ISIS) mounted on the 4.2m William Herschel Telescope (WHT). }
   {As predicted by theoretical models,  we find a higher degree of polarization in the R band when compared to polarization in the I band for  3/4 of these ultra cool dwarfs. This suggests that dust scattering asymmetry is caused by oblateness . We also show how these measurements fit the theoretical  models. A case for variability of linear polarization is found, which suggests the presence of randomly distributed dust clouds. We also discuss one case for the presence of a cold debris disk.
  }
   {}

   \authorrunning{R. Tata et al.}            
   \maketitle
%

\section{Introduction}

A large number of ultra cool dwarfs have been detected in the last decade, and our understanding of these faint objects has kept improving. One of the challenging and fundamental aspects in the study of these objects is to understand the properties and distribution of condensate dust in the atmosphere. Observations of L dwarfs with effective temperatures of 1400-2200 K have led to the investigation of dust condensates in their atmospheres (\cite {Kirkpatrick}; \cite {Tsuji1996}). Because of complete gravitational settling, grains are expected to condense beyond the visible atmosphere for objects with effective temperatures below 1400 K (T-Dwarfs -  \cite{Allard 2001}; \cite{Chabrier00}). At higher effective temperatures (1400-2200 K), grains can be present in the visible atmosphere because of incomplete gravitational settling (\cite {Burrows and Sharp}; \cite {Burrows 2001}; \cite {Ackerman 2001}; \cite {Allard 2001};  \cite{Tsuji 2004};   \cite{co03}; \cite{helling}  ). Recent discoveries of blue L dwarfs and L-T transition type dwarfs (as identified in \cite {Knapp  2004}; \cite {Chiu 2006}; \cite{Tsuji03}) have brought forth models which could explain this phenomenon (eg. \cite {BSH 2006}, \cite {Knapp  2004}) by mechanisms which involve dust settling. It would be very important to validate these mechanisms.

Linear polarization could be a very useful tool in understanding the  observationally poorly constrained dust properties in the atmospheres of L dwarfs. The possibility of detecting polarization at optical wavelengths from grains in the atmospheres of L dwarfs was first raised by \cite{SKR 2001}. Fast rotation of L dwarfs  will induce the shape of their photosphere into the form of an oblate ellipsoid,(\cite {Basri 2000}) and this nonsphericity will lead to the incomplete cancellation of the polarization from different areas of the stellar surface (\cite {SKR 2001}). This prediction was first confirmed by the detection of linear polarization at 768 nm from a few L dwarfs by \cite {Menard 2002}. Recently, \cite{Zap 2005} have reported R and I band detection of linear polarization from several L dwarfs. Since polarization in the optical is unlikely to be due to Zeeman splitting of atomic or molecular lines or by synchrotron radiation, the observed  polarization can be explained by single dust scattering in a rotationally induced oblate atmosphere (\cite {S 2003}; \cite{SK 2005}) or it could be due to large and randomly distributed dust clouds (\cite{Menard 2002}).

In this paper, we report polarization measurements of  3 L dwarfs (L0-L5) and one M9.5 dwarf  with WHT/ISIS in both I and R bands .  We also discuss our results comparing them with the recently published results of \cite{goldman09}.

 Our measurements show the general trend that  polarization is higher in the R band than  the one in the I band. This trend strongly supports the presence of dust in the atmosphere of L dwarfs as it is very  unlikely that any other mechanisms (such as the presence of magnetic field) can explain this observation   at optical wavelengths (\cite{Menard 2002}). We also discuss how the theoretical models  (see Sect. 3)  successfully fit our measured data.

   

\section{Target selection and observations}
Four very nearby (7pc -15pc)  ultra cool dwarfs were selected (SpT M9.5 - L5) as they are among the brightest and nearest ultra cool dwarfs with no known infrared excess and no evidence of multiplicity(see Table 1).
These selection criteria ensure that the targets are bright enough sources in the R band to get high S/N and to avoid  other than intrinsic sources of polarization such as circumstellar disks or multiplicity.
For calibration, one polarized (Cyg OB2 A Whittet et.al. 1992) and one unpolarized standard  star were observed at two different times during the night. All the objects were observed in a way that they insured they were acquired at the same position on the detector (5 pixel box).  This procedure minimized contamination caused by instrumental polarization within the detector and variations in  the optical path.\\

The polarimetric observations were obtained using the Intermediate dispersion Spectrograph and Imaging System (ISIS) which is mounted at the Cassegrain focus of the 4.2m William Herschel Telescope (located in La Palma, Canary Islands, Spain).
ISIS in polarization mode is a modulation polarimeter with a double-beam analyzer (the calcite plate) and a rotating halfwave plate modulator. ISIS is equipped with two detectors: a blue-sensitive EEV12 (4096x2048 pixels) and a red-sensitive RED+ (4096x2048) detector. In our program, we have used the RED+ detector.\\

Images were obtained using Bessel R- and I- filters  centered on 641 and 812 nm, respectively, on June 18, 2006 (UT Date). The night was photometric  with stable average seeing of 1.0 arcsec .

 The raw images were bias-subtracted and flat-fielded before performing aperture photometry. The flat-field images were obtained with the polarimeter optics.
Fluxes were obtained for 0.8, 1.0, 1.2, 1.5, 2.0 times the average FWHM for each object. The best aperture was chosen to be 1.5 times FWHM based on minimum photon contribution of nearby sources, variable sky contribution, and maximum signal-to-noise ratio of the measurements. The average FWHM of all images was 4.0 pixels which corresponds to 1.0 arcsec.  
We have only one set of measurements for each object. Therefore, we have estimated the uncertainty in the degree of polarization from various apertures (a similar method was used by \cite {Zap 2005} for some of the objects).
There was no significant instrumental polarization found as the unpolarized standard measured D(p)=0.086\% $\pm$ 0.002 .


\section{Theoretical modeling of polarization}

 Polarization is a measure of anisotropy in the radiation field and is caused by either scattering 
or is due to the presence of magnetic field. The state of the polarization of light is described by the Stokes parameters, {\it I, Q, U} and {\it V}.
The parameter {\it  I} is the total scalar specific intensity of radiation.
It is the complete flux of radiant energy inside the unit intervals of
frequency, time, solid angle, and area perpendicular to the flux.
This flux includes all radiation independently of polarization.
Polarization is described by the parameters {\it Q, U, V}. These parameters
are proportional to the scalar specific intensity and have the same dimension.
 Q and U represent the linearly polarized component, and V represents the circularly polarized component.
For linear polarization, {\it V=0} and the degree of polarization is given as
$p=\sqrt{Q^2+U^2}/I$. If we consider axial symmetry, then $U=0$ and in that
case we define the degree of polarization $p=-Q/I$. The sign convention
is such that if $p>0$, the light is polarized perpendicular to the scattering
plane, and if $p<0$, the light is polarized parallel to the scattering plane.
 For an unresolved stellar object, the Stokes parameters are
integrated over the stellar disk.

From our obtained images, the degree of polarization and  the polarization angles are calculated using the following equations:
   \begin{equation}
     R_Q^2 = \frac{\textit{o}(0^\circ)/\textit{e}(0^\circ)}{\textit{o}(45^\circ)/\textit{e}(45^\circ)}\, ;   
       Q/I = \frac{R_Q - 1}{R_Q + 1}\ 
    \end{equation}
    \begin{equation}    
       R_U^2 = \frac{\textit{o}(22.5^\circ)/\textit{e}(22.5^\circ)}{\textit{o}(67.5^\circ)/\textit{e}(67.5^\circ)}\, ;   
       U/I = \frac{R_U - 1}{R_U + 1}\
   \end{equation}
   \begin{equation}    
       p = \sqrt{(Q/I)^2 + (U/I)^2 }
    \end{equation}
    \begin{equation}    
       \theta = 0.5  tan^{-1} { \frac{U/I }{Q/I}\ }
    \end{equation}
   
   where 
   
   \[
      \begin{array}{lp{0.8\linewidth}}
       o, e & ordinary  and extraordinary fluxes respectively of the dual images of the program source on a 	single frame \\ 
         p  & degree of polarization\\
         \theta & polarization angle\\
        I &  the total scalar specific intensity of radiation\\
	Q/I, U/I & normalized Stokes parameters
      \end{array}
   \]

As pointed out by \cite{Menard 2002}, the observed linear polarization in the optical
cannot be due to magnetic field, and scattering remains the most viable mechanism
for yielding the detected linear polarization. Polarimetric observation
at the R and I bands by \cite{Zap 2005} shows that polarization decreases significantly 
with the increase in wavelength, which strongly supports  this argument (\cite{SK 2005}). In the present 
investigation, we report detection of polarization at both R and I bands which shows
the same wavelength dependency and hence strengthens the case for scattering polarization.     
 If the dust density is low, then the single scattering approximation is reasonable for the
region where the dust optical depth $\tau_d < 1$ because
scattering by atoms and molecules does not significantly contribute to polarization.
However, in the optical region $\tau_d$ is much higher than 1, and therefore multiple
scattering is the appropriate process.
Multiple scattering reduces the amount of polarization significantly because
the planes of scattering are oriented randomly and cancel each others' contributions.
In the present paper, however , we adopt a simple single scattering polarization
model because the main aim is to explain the wavelength dependency of the observed
polarization. Nevertheless, we find that a single dust scattering model in  rotation-induced oblate L dwarfs can fit the observed data well within the permissible range
of parameter space of dust composition, size, geometrical distribution and rotation
period.  A detailed modeling with multiple scattering is in progress. 

  The simple theoretical model adopted here to explain the observed polarization is
described in details in \cite{SK 2005}.  At an edge-on view, the degree of polarization
integrated over the stellar disk is given by : 
{\tiny
\begin{eqnarray}\label{pol}
p(\lambda)=\frac{\lambda^2}{g}\int^{P_{bot}}_{P_{top}}\frac{n(P)dP}{\rho(P)} 
\sum^{\infty}_{l=2}
\left\{\alpha^2(l,m)P_l^m(0)F_{l2}\int^1_{-1}\frac{P_l(\mu)}{[1+(A^2-1)\mu^2]^{1/2}}d\mu
\right\}
\end{eqnarray}
}
In the above expression, $P$ is the total pressure (gas plus dust), 
$P_{top}$ and $P_{bot}$ are the pressures at the top (deck) and the bottom (base) of
the dust layer, $n(P)$ is the grain number density at different pressure heights, $\rho(P)$
is the total density at any pressure level, and $A$ is the ratio of the equatorial radius
to the polar radius. Since the dust density is small  compared to the gas density,
$\rho(P)$ is the effective gas density and is equal to $nRT/P$ where $n$ is the mean
molecular weight and $R$ is the gas constant. In the above equation, $P_l(\mu)$ is the
Lengendre polynomial where $\mu=\cos\theta$, $\theta$ being the scattering angle
and $F_l2$ is given by
\begin{equation}
F_{lm}=\alpha(l,m)\int^1_{-1}\frac{i_1-i_2}{2}P^m_l(\cos\theta)d(\cos\theta).
\label{abc2a}
\end{equation}
where 
\begin{equation}
\alpha(l,m)=\left[\frac{(2l+1)(l-m)\!}{4\pi(l+m)\!}\right]^{1/2},
\end{equation}
and $P^m_l$ is the associated Legendre function of the first kind.
$i_1$ and $i_2$ are the scattering functions given by \cite{van57}.

The vertical dust distribution and the location of the cloud base and deck in the atmosphere
are calculated based on the one dimensional heterogeneous cloud model of \cite{co03}.
This model assumes chemical equilibrium throughout the atmosphere and uniform
density distribution across the surface of an object at each given pressure
and temperature.
 The number density of cloud particles in this model is given by

\begin{eqnarray}\label{density}
n(P)=q_c \left( \frac{\rho}{\rho_d}\right) 
\left(\frac{\mu_d}{\mu}\right)
\left(\frac{3}{4\pi a^3}\right),
\end{eqnarray}
where $\rho$ is the mass density of the surrounding gas, $a$ is the cloud
particle radius, $\rho_{d}$ is the mass density of the dust condensates,
$\mu$ and $\mu_d$ are the mean molecular
weight of atmospheric gas and condensates respectively.
The condensate mixing number ratio ($q_c$) is given as

\begin{eqnarray}
q_c=q_{below}\frac{P_{c,l}}{P}
\end{eqnarray}
for heterogeneously condensing clouds.
In the above equation, $q_{below}$ is the fraction of condensible vapor just
below the cloud base, $P_{c,l}$ is the pressure at the condensation
point. 

A log-normal size distribution is adopted for the spherical grains (\cite{marley01}).
The formation of dust makes it a prohibitive task to develop a fully consistent
atmospheric model for ultra-cool dwarfs. This is mainly because of the fact
that the presence of dust clouds affects the radiative equilibrium of the
upper atmosphere and hence alters the T-P profile from that of a cloud-free
atmosphere. On the other hand, the T-P profile dictates the position and
the chemical equilibrium of condensates. Allard et al. (2001) presented
atmospheric models for  two of the  limiting cases, e. g., one with inefficient
gravitational settling wherein the dust is distributed according to chemical
equilibrium predictions (AMES-dusty) and another with efficient gravitational
settling in which situation dust has no effect on the thermal structure
(AMES--cond). \cite{Tsuji 2004} have proposed a Unified Cloudy Model (UCM)
in which the segregation of dust from the gaseous mixture takes place in all
the ultra-cool dwarfs and at about the same critical temperature.
\cite{marley01} treat the upward convective mixing of a gas, its condensation
and the sedimentation of the condensate through the atmosphere of the object while
 \cite{woi04} consider an ensamble of dust grains falling downwards from the top
of the atmosphere. A detailed comparison
of different atmospheric models of L dwarfs is presented in \cite{helling}.

The oblateness of a rotating object has been discussed by \cite{cha33}
in the context of
polytropic gas configuration under hydrostatic equilibrium.  For a slow rotator,
the relationship for the oblateness $f$ of a stable
polytropic gas configuration under hydrostatic equilibrium is given by
\begin{eqnarray}\label{obl}
f=\frac{2}{3}C\frac{\Omega^2R^3_e}{GM},
\end{eqnarray}
where $M$ is the total mass, $R_e$ is the equatorial radius, and $\Omega$ is
the angular velocity of the object which is related to the linear velocity $V=
\Omega\times R_e$. $C$ is a constant whose value depends on the polytropic index.
For a polytropic index of $n=1.0$, $C=1.1399$, which is appropriate for
Jupiter (\cite{hub84}). For non-relativistic completely degenerate gas,
$n=1.5$ and $C=0.9669$.
  
 The effective temperature of the L dwarfs of different
spectral type is determined by adopting a sixth order polynomial fit given by
\cite{goli} which is based on  bolometric luminosities.
The $T_{eff}$ calibration of \cite{goli} agrees well
in the interval L3-L8, but there are significant differences in earlier types.
In our calculations for the degree of polarization, the effective
temperature $T_{eff}$ is used and hence the degree of polarization should
be considered strictly as a function of $T_{eff}$ rather than of spectral
type.  The mass and radius
of the L dwarfs of different spectral types are estimated by adopting the empirical
relationship given by \cite{marley}.
  
\begin{table*}
\caption{Target list}             
\label{table:1}      

\centering                          
\begin{tabular}{l l l l l l l l l l r}        
\hline\hline                 
Object  & RA & DEC &  $SpT$&R&I&J&d& H$\alpha$ & ref  \\ 
&&&&(mag)&(mag)&(mag)&(pc)&EW($\AA$)&\\
\hline                        
2MASSW J1438082+640836 & 	14 38 08.2   & 	+64 08 36 & 	M9.5 & \ldots	 & 	16.8 & 	12.92 & 	7.2 $\pm$0.5&$-2.36$  & 1 \\
2MASSW J1507476-162738 & 	15 07 47.69   & 	-16 27 38.6 & 	L5 & 	18.9 & 	16.5 & 	12.8 & 	7.33  $\pm$ 0.03&$-5.01$ & 	3,4,5   \\  
2MASS J17312974+2721233 & 	17 31 29.74   & 	+27 21 23.3 & 	L0 & 	\ldots &\ldots 	 & 	12.09 & 	11.80 $\pm$ 0.70 &$-5.98$ &1,2  \\
2MASSI J1807159+501531 & 	18 07 15.93   & 	+50 15 31.6 & 	L1.5 & \ldots	 & 	\ldots & 	12.96 & 	14.6  $\pm$ 1.0 &$-1.38$ &1,2,3 \\

\hline                                   
\end{tabular}
\note { ref- 1. \cite{SCH 2007},  2. \cite{Jameson 2007},  3. \cite{Cruz 2003},  4. \cite{Reid 2000}, 5. \cite{Knapp 2004} \\
H$\alpha$ EW from \cite{SCH 2007}}
\end{table*}   

\section{Results}

 We find a trend (more data are required to confirm our theory) of higher polarization in the R band when compared
to the I band. This wavelength dependency strongly supports the argument by \cite{SK 2005} that the
polarization arises due to scattering and not because of magnetic field. In dust scattering as described
by Mie theory, the amount of polarization depends on the ratio of the grain radius to the
wavelength. For the same kind of dust species, the polarization usually peaks when the ratio is
one. As a consequence, the increase in polarization with the decrease in wavelength implies the
presence of sub-micron size grains in the photosphere of the L dwarfs. We present our measurements in Table 2, and our model fit in Figure~1. 
 One of the L dwarfs from \cite{Zap 2005} (2MASSW J1507476-162738) shows a
null polarization in our measurements in the I band, whereas \cite{Zap 2005}
present a higher polarization (1.36$\%$ $\pm$0.30) for the same object. 
This suggests variability in linear polarization which in turn suggests
atmospheric activities  like dynamical variations of the cloud cover.

 We also find relatively high polarization in the L0 dwarf. Additionally, from \cite {SCH 2007},
the H$\alpha$ equivalent width is the highest for the L0 dwarf among the objects from our sample.
This could be an indirect evidence of a disk around this ultra cool dwarf.  We therefore searched the {\it Spitzer} public archive
for mid-IR data. 2MASS J17312974+2721233 has been observed with IRAC and IRS in the course of program 3136 (P.I. Cruz), and we retrieved the pipeline processed data. We extracted the IRAC photometry using standard PSF photometry procedures within the Interactive Data Language. Uncertainties were estimated from the Poisson noise weighted by the coverage maps of the mosaics. Table~\ref{irac} gives a summary of the photometry. Figure~\ref{sed} shows the spectral energy distribution (SED) of the source and a L0 comparison object from the literature (2MASS~J1204+3212, Patten et al., 2006). 2MASS J17312974+2721233 does not show any significant mid-IR excess up to 15~$\mu$m. The presence of a young circumstellar disc can therefore be ruled out at a high level of confidence. In the current state of the data, we cannot rule out the presence of a cold debris disc, as it would produce an excess at longer wavelengths.

\begin{table*}
\caption{Polarization measurements}             
\label{table:2}      
\centering                          
\begin{tabular}{l l l l l l }        
\hline\hline                 
Object  &$SpT$   &Filter &  Exposure Time &  D(p) & $\theta$  \\    
 &   & &  (seconds) &  (\%)&  (degrees)  \\ 
\hline                        
2MASSW J1438082+640836 & M9.5 &   I  & 	90    & 	0.482 $\pm$0.071 & 66.24  \\
 &  & R & 300  & 0.54  $\pm$0.097 & 115\\
 \hline
 2MASSW J1507476-162738 & L5  & 	I  & 150 & 	0.0 $\pm$0.036 & \ldots   \\  
 &  & R &300 & 	0.216 $\pm$0.022 & 100.26   \\
 \hline
2MASS J17312974+2721233  &  L0  & 	I  &  60  & 	5.195 $\pm$ 0.854 &  52.57   \\
 &  & R & 240 & 0.666 $\pm$ 0.169 & 111.08\\
 \hline
2MASSI J1807159+501531 & 	L1.5 & 	I & 60  & 	0.711 $\pm$ 0.142  & 46.67 \\
 &  &  R & 120 & 1.669  $\pm$ 0.605& 101.2 \\

\hline                                   
\end{tabular}
\note { The uncertainty in $\theta$ due to calibration errors is about 1.2 degrees}
\end{table*}

%
   \begin{figure*}
\centering
   \includegraphics[width=\textwidth]{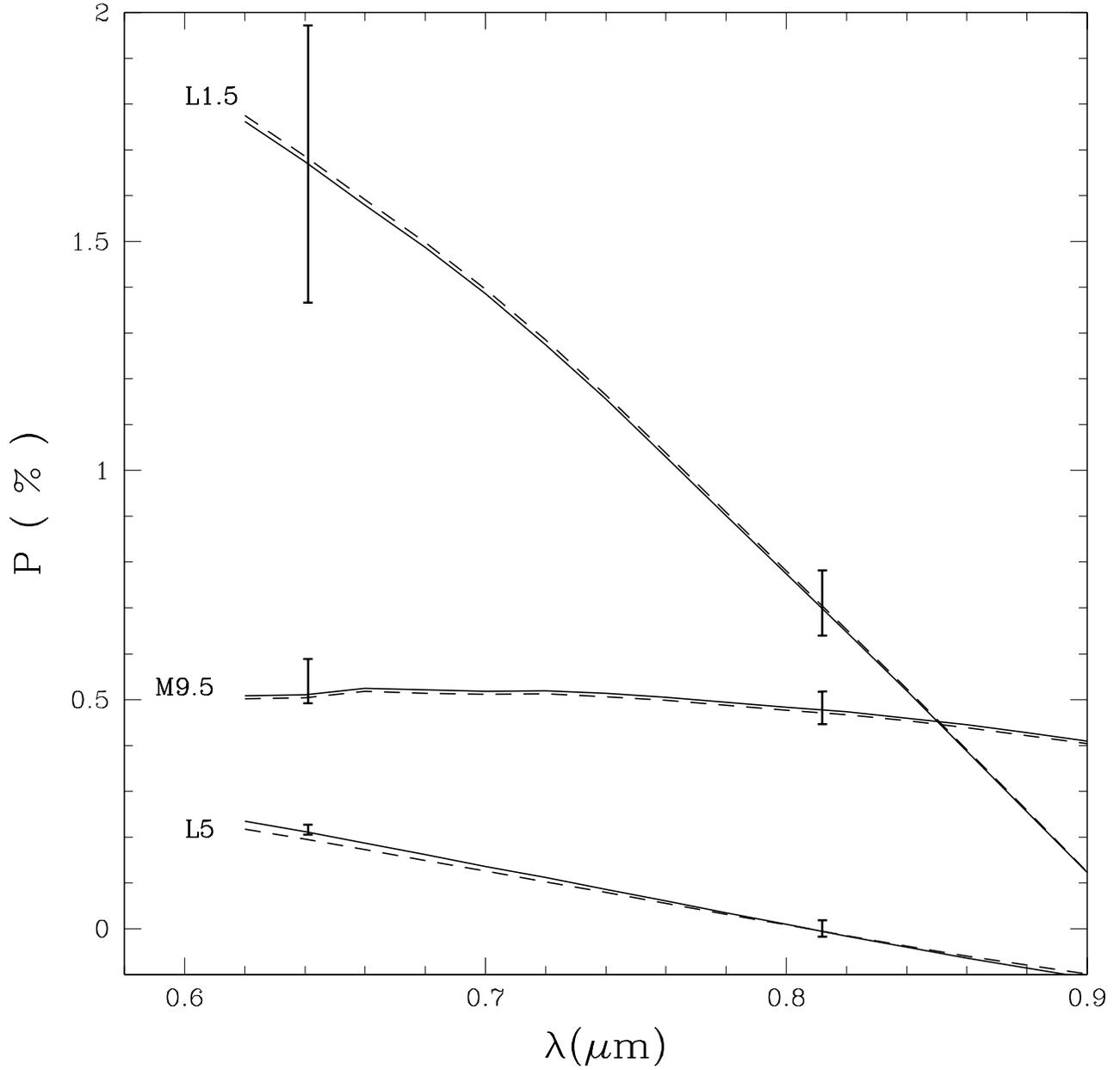}
      \caption{ Best model fit of the observed data. The solid lines
represent the model with the polytropic index n=1.0 and dashed lines
represent that with n=1.5. For other parameters see Table 3.
  }
         \label{Figmodelfit}
   \end{figure*}

   
%

\begin{table*}
\caption{Model fit}             
\label{table:3}      
\centering                          
\begin{tabular}{l l l l l l }        
\hline\hline                 
Object  &$SpT$ & n  & $d_0 $ & log(g) & V  \\    
  & &  &(micron) & (cgs) & (km s$^{-1}$)  \\    
\hline                        
2MASSW J1438082+640836 & M9.5 & 1.0 & 0.7 & 	5.0    &  18.7 \\
 &  & 1.5& 0.7 & 	5.0    &  20.5 \\
 \hline
2MASSW J1507476-162738 & L5  & 1.0 & 0.44 & 	5.255    &  27.2 \\
 &  & 1.5& 0.44 & 	5.230    &  27.2 \\
 \hline
2MASSI J1807159+501531 & 	L1.5 & 1.0 & 0.5 & 	5.41    & 76.0 \\
 &  & 1.5& 0.44 & 	5.37   &  76.0 \\
 \hline

\hline                                   
\end{tabular}
\note { Bailer-Jones 2004 has measured the projected rotational velocity for
2MASSW J1507476-162738 as 27.2 km s$^{-1}$, while Reiner \& Basri 2008 have reported the projected velocity for 2MASSW J1807159+501531 to be 76 km s$^{-1}$.} 
\end{table*}

\begin{table*}
\caption{IRAC photometry of 2MASS J17312974+2721233}             
\label{irac}      
\centering                          
\begin{tabular}{l c}        
\hline\hline                 
Wavelength  & Flux \\
(micron) & (mJy) \\
\hline                        
3.6 & 21.82$\pm$0.02 \\
4.5 & 14.16$\pm$0.02 \\
5.8 & 11.30$\pm$0.06 \\
8.0 & 6.91$\pm$0.03 \\
\hline                                   
\end{tabular}
\end{table*}

   \begin{figure*}
\centering
 \includegraphics[width=\textwidth]{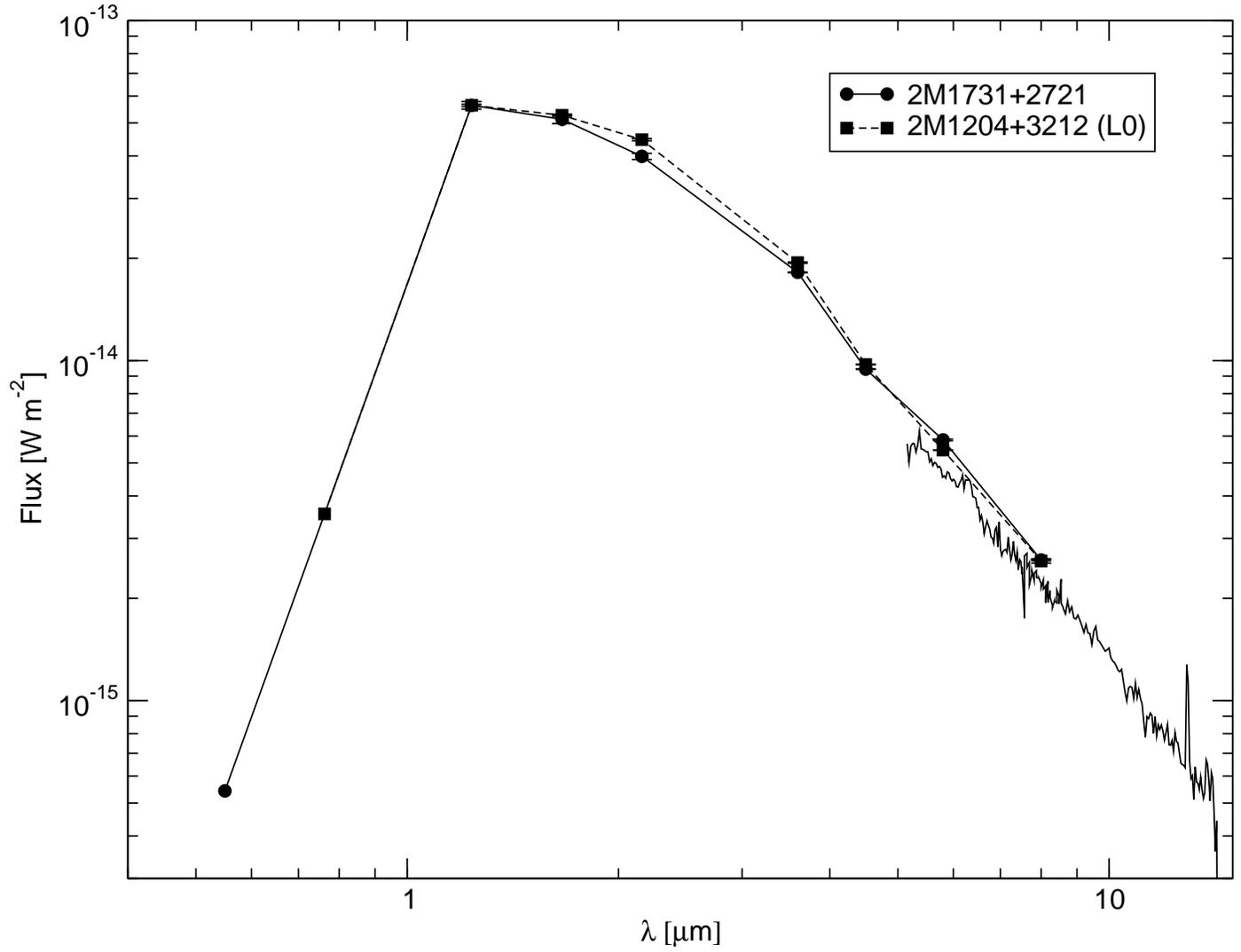}
      \caption{Spectral energy distribution of 2MASS J17312974+2721233 (dots). V-band photometry from the  LSPM-North proper-motion catalog of nearby stars (L\'epine et a., 2005). J,H and Ks photometry from 2MASS. IRAC and IRS photometry from this work. A comparison of L0 dwarf (2MASS~J1204+3212, Patten et al., 2006) scaled to the same J-band flux is overplotted (squares). The target does not show any significant excess compared to the field L0 dwarf.
  }
         \label{sed}
   \end{figure*}

  The surface gravity of L dwarfs older than approximately a few
hundred million years varies from $g=10^5$ to $3\times10^5$ $cm s^{-2}$. 
Evolutionary models by \cite{Chabrier00} show younger L dwarfs to have a surface gravity
smaller than $10^5$ $cm s^{-2}$. Our best fit model parameters are presented in Table~3  (best fit obtained by visual inspection).
For the L dwarfs 2MASSW J1438082+640836,
we find the best fit with log(g)=5.0, and rotational velocity V=18.7 km s$^{-1}$ when $n=1.0$, and
V=20.5 km s$^{-1}$ when $n=1.5$.
For a fixed rotational velocity, the oblateness is lower when the polytropic
index $n$ is higher. As a result, higher rotational velocity is needed to fit the observed data
if $n=1.5$. The projected rotational velocity of the L dwarf 2MASSW J1507476-162738 is measured
by \cite{B-J 2004} while the projected rotational velocity of 2MASSW J1807159+501531 is estimated by \cite{RNB 2008}. In the absence of any knowledge on the
projection angle, we consider the minimum rotational velocity of these objects as V=27.2 km s$^{-1}$ and V=76 km s$^{-1}$.
We find the best fit for the observed polarization of 2MASSWW J1507476-162738 
with  Log(g)=5.255 for $n=1.0$ and Log(g)=5.230 for $n=1.5$
 For 2MASSW J1807159+501531 the best fit is obtained with log(g)=5.41,
when $n=1.0$, and log(g)=5.37 when $n=1.5$. 
Note that the oblateness decreases with the increase in surface gravity.
 For all the cases, the observed polarization
profiles can be fitted with sub-micron size grains, and the mean size of grains that are
required to fit the observation is consistent with the recent theoretical calculations
of dust properties (\cite{woi04}; \cite{woi03}).   
 Polarization measurements for one of the above three objects (2MASSW J1507476-162738) were also recently published \cite{goldman09}. Our results are consistent with the \cite{goldman09} measurements within 1$\sigma$ error bars for this object. This work differs from \cite{goldman09} in the aspect that our sample could reproduce the predictions of \cite{SK 2005} whereas the \cite{goldman09} sample does not.   Both the studies (this paper and \cite{goldman09}) have sparse statistics, which discourages us to make any speculations on this apparent discrepancy.

\section{Conclusions}
   \begin{enumerate}
      \item We report linear polarizaion measurements of 4 very nearby ultra cool dwarfs in the R and I bands.
      \item We find that there is a trend (3 out of 4) of a higher degree of  polarization at shorter wavelengths (R band) when compared to the I band as predicted by the theoretical models of \cite{SK 2005}.
      \item  The L0 dwarf 2MASS J17312974+2721233 is interesting because of its relatively high polarization and requires follow-up studies.
      \item We also fit theoretical models to predict the dust grain size and rotational velocities of three of the ultra cool dwarfs.
      \item We find evidence for variability in the linear polarization for (2MASSW J1507476-162738). This suggests atmospheric activities like dynamical variations of the cloud cover in this object.
   \end{enumerate}

\begin{acknowledgements}
     This work was partially funded by the Spanish MICINN under the Consolider-Ingenio 2010 Program grant CSD2006-00070: First Science with the GTC  (http://www.iac.es/consolider-ingenio-gtc)
 \end{acknowledgements}


\begin{thebibliography}{}

   \bibitem[Ackerman \& Marley 2001]{Ackerman 2001} Ackerman \& Marley 2001,
   ApJ, 556, 872.

   \bibitem[Allard et al. 2001]{Allard 2001} Allard et al. 2001,
   ApJ, 556, 357.
   
   \bibitem[Bailer-Jones 2004]{B-J 2004} Bailer-Jones, C. A. L 2004,
   A \& A, 419, 703. 
   
   \bibitem[Basri et al.  2000]{Basri 2000}Basri et al.  2000,
   ApJ, 538, 363.
   
    \bibitem[Berger et al. 2005]{Berger 2005}Berger et al.  2005,
   ApJ,627, 2, pp. 960-97
   
   \bibitem[Burrows et al. 2001]{Burrows 2001} Burrows et al. 2001,
   RvMP, 73, 719.    
   
   \bibitem[Burrows \& Sharp 1999]{Burrows and Sharp} Burrows \& Sharp 1999,
   ApJ, 512, 843.
   
   \bibitem[Burrows, Sudarsky \& Hubeny 2006]{BSH 2006} Burrows, Sudarsky \& Hubeny 2006, ApJ, 640, 1063.

   \bibitem[Chabrier et al. 2000]{Chabrier00} Chabrier,G.  et al. 2000, ApJ, 542, 464

   \bibitem[Chandrasekhar 2003]{cha33} Chandrasekhar, S. 1933, MNRAS, 93, 539
 
   \bibitem[Chiu et al. 2006]{Chiu 2006} Chiu et al. 2006,
   AJ, 131, 2722.
   
   \bibitem[Cooper et al. 2003]{co03} Cooper, C. S., Sudarsky, D., Milsom, J. A.,
             Lunine, J. I. \& Burrows, A.  2003, ApJ, 586, 1320. 
             
   \bibitem[Cruz et al. 2003]{Cruz 2003} Cruz et al. 2003, 
   AJ, 126, 2421.      

  \bibitem[Goldman et al. 2009]{goldman09} Goldman,B  et al. 2009, A\&A, 502,3, 929.

   \bibitem[Golimowski et al. 2004]{goli} Golimowski, D. A., et al. 2004, AJ, 127, 3516. 

   \bibitem[Helling et al. 2008]{helling} Helling, Ch. et al. 2008, MNRAS, 391, 1854. 

   \bibitem[Helling 2003]{helling03} Helling, Ch. 2003, Rev. Modern Astron. 16, 15. 

   \bibitem[Hubbard 1984]{hub84} Hubbard, W. B. 1984, Planetary Interiors
      (New York;Van Nostrand Reinhold).

   \bibitem[Jameson et al. 2007]{Jameson 2007} Jameson et al. 2007,
   AJ, 119, 339.

   
   \bibitem[Kirkpatrick et. al. 1999]{Kirkpatrick} Kirkpatrick et al.1999,
        ApJ 519 , 834.
     
   \bibitem[Knapp et al. 2004 ]{Knapp 2004} Knapp et al. 2004,
   AJ, 127, 3553. 
   
   \bibitem[Marley et al. 1996]{marley} Marley, M. S., et al. 1996, Science, 272, 1919.

   \bibitem[Ackerman \& Marley 2001]{marley01} Ackerman, A. S. \& Marley, M. S., 2001,
    ApJ, 556, 872.

   \bibitem[Menard et al. 2002]{Menard 2002} Menard et al. 2002,   A \& A, 396L,35.     

   \bibitem[Patten et al. 2006]{Patten 2006} Patten et al. 2006,  ApJ, 651, 502.     
   
   \bibitem[Reid et al. 2000]{Reid 2000} Reid et al. 2000. AJ, 119, 339.

   \bibitem[Reiners \& Basri 2008]{RNB 2008} Reiners \& Basri 2008, ApJ 684, 1390.    

   \bibitem[Schmidt et al. 2007]{SCH 2007} Schmidt et al. 2007,
   AJ, 133, 2258.
         
   \bibitem[Sengupta 2003]{S 2003}Sengupta 2003,
    ApJ, 585, L155.
    
   \bibitem[Sengupta \& Krishan 2001]{SKR 2001} Sengupta \& Krishan 2001,
   ApJ,561L,123.
   
   \bibitem[Sengupta \& Kwok  2005]{SK 2005} Sengupta \& Kwok, 2005,
   ApJ, 625, 996.
 
   \bibitem[Tsuji et al. 1996]{Tsuji1996} Tsuji et al.1996,
   A\&A 308, L29-L32
   
    \bibitem[Tsuji \& Nakajima 2003]{Tsuji03}  Tsuji \& Nakajima 2003,
   ApL 585,2, L151-L154
       
   \bibitem[Tsuji et al. 2004]{Tsuji 2004} Tsuji et al. 2004,
   ApJ, 607, 511.

   \bibitem[van de Hulst 1957]{van57}
    van de Hulst, H. C. 1957, Light Scattering by Small Particles (New York;
    Willey)

   \bibitem[Woitke \& Helling 2004]{woi04} Woitke, P., \& Helling, Ch. 2004, A\&A, 414, 335
   \bibitem[Woitke \& Helling 2003]{woi03} Woitke, P., \& Helling, Ch. 2003, A\&A, 399, 297

   \bibitem[Zapatero Osorio et al. 2005]{Zap 2005} Zapatero Osorio et al. 2005,
   ApJ, 621, 445.
   
   
   \end{thebibliography}
\end{document}